\documentclass[a4paper,10pt]{article}

\usepackage{amsmath}

\title{Polyakov Loops for the ABJ Theory }

\author{ Mir Faizal$^1$ and Tsou Sheung Tsun$^2$ \\
$^1$Department of Physics and Astronomy, \\  University of Waterloo,   Waterloo,\\
Ontario N2L 3G1, Canada\\
$^2$Mathematical Institute, University of Oxford
\\ Oxford
OX1 3LB, United Kingdom 
 }

\date{}

\begin{document}

\maketitle

\begin{abstract}

In this paper we will first construct Polyakov loops for the ABJ theory. 
Then we will construct the connection and curvature in this loop space. 
We will also analyse certain  generalization of Polyakov loops and apply them 
to the ABJ theory. Finally, we will use this formalism for analysing topological defects like 
monopoles in the ABJ theory. 

\end{abstract}

\section{Introduction}

According to the $AdS/CFT$ correspondence  
the eleven dimensional supergravity  on $AdS_4 \times S_7$ is dual to a superconformal field theory
describing multiple M2-branes. 
This superconformal field theory  has to have 
$\mathcal{N} = 8$ supersymmetry. This is because apart from a 
 constant closed 7-form on $S^7$,
$AdS_4 \times S^7 \sim [SO(2,3)/ SO (1, 3)]\times  [SO(8)/ SO(7)] \subset OSp(8|4)/[SO(1,3) \times SO(7)]$. The 
 $OSp(8|4)$ gets realized as $\mathcal{N} = 8$ supersymmetry of the dual superconformal field theory.
 The transverse coordinates of the M2-branes give rise to eight gauge valued scalar fields. 
 Apart from these eight gauge valued scalar fields, this theory also has  sixteen physical 
fermions.  The gauge fields of this theory do not have any on-shell degrees of freedom. 
 A  theory called the BLG theory satisfies these properties 
 \cite{1b, 2b, 3b, 4b, 5b}. However, the gauge symmetry of the BLG theory is based on a 
Lie 3-algebra and the only known example of a Lie 3-algebra is $SO(4) \sim SU(2) \times SU(2)$. 
So, the BLG theory can only describe two M2-branes.

It has been possible to generalize the BLG theory to a superconformal field theory 
describing any number of M2-branes on 
$AdS_4 \times S_7/ Z_k$ 
\cite{abjm, ab, ab1, ab0}. This theory called the ABJM theory only has $\mathcal{N} =6$ supersymmetry 
and $SO(6)$ $R$-symmetry. However, as it considers with the BLG theory for two M2-branes, 
it is expected that  its supersymmetry would get enhanced to 
full $\mathcal{N} = 8$ supersymmetry.
In fact, the supersymmetry for the ABJM can gets enhanced to $\mathcal{N} =8$ supersymmetry 
for Chern-Simons levels, $k = 1, 2$, by the use of monopole operators \cite{mp, mp1, mp2, mp0}.
In the ABJM theory  the matter fields are in the bi-fundamental representation 
of the gauge group $U(N) \times U(N)$ and the double gauge fields are in the 
adjoint representation. 
A further generalization of the ABJM theory to  a theory describing fractional M2-branes has 
been made \cite{5bc, 5ba, 5a1, 5a2, 5a}. This theory is called the ABJ theory 
and in it the gauge fields are described by the gauge group 
$U(M) \times U(N)$ with $M \neq N$ \cite{1, 2}. The matter fields are again in the bi-fundamental representation 
of this gauge group and the double gauge fields are in the adjoint representation.

Wilson loops for the ABJ theory have been studied and they are given by
semi-classical string surfaces in the dual string theory picture \cite{3, 4}. 
The most symmetric string of this kind preserves half the supersymmetry. The 
dual field theory operator to it has also been  constructed  using a superconnection \cite{5}. 
In this superconnection, 
the scalar fields occur in bi-linears combinations
and the fermions appear linearly. Thus, the fermions transform in the bi-fundamental representation and 
appear in the off-diagonal block. 
The bi-linear product of the scalars transforms in the adjoint representations. 
So, the scalars appear in the diagonal blocks along with the gauge fields. 
The fermions couple to  Grassmann even quantities and thus the off-diagonal blocks 
contain Grassman odd quantities. 
It may be noted that Wilsons loops which preserve $1/6$ of the total supersymmetry 
have also been studied \cite{6, 7}. In fact, a matrix model corresponding to the 
vacuum expectation value for the $1/6$ BPS Wilsons loop has been constructed \cite{7a}.

In this paper we introduce Polyakov loops as the variables to be
used.  In mathematical language these are the holonomies of closed
loops in space-time, and they are sometimes also called Dirac phase factors
in the physics literature.  Although they are defined via parametrized
loops in space-time, they are independent of the parametrization
chosen.  They are therefore gauge group-valued functions of the
infinite-dimensional loop space.
The main difference between a Polyakov loop and a Wilson loop is that in the Wilson loop a trace is taken and 
no such such trace is taken in the Polyakov loop \cite{p1}. 
In this paper we will study the Polyakov loops for the ABJ theory.  
Polyakov loops have been used for deriving a duality in non-abelian gauge theories 
\cite{pq1, pq2}.
This duality has been used for analysing the  't Hooft's  order-disorder parameters \cite{1p}.
A   Dualized Standard Model 
has also been constructed using this duality \cite{pq4, p9}.
In this  model  three 
generations of fermions are produced by the breaking of a  dual color $SU(3)$  symmetry \cite{pq0, q01}. 
The resulting scheme give a method for calculating a 
fermion mass hierarchy along with the 
mixing parameters of the Standard Model fermions \cite{qp1, pq01}. 
Dual Feynman Rules for Yang-Mills theories with a monopole have also  been 
analysed using Polyakov loops \cite{pq5}. Polyakov loops for supersymmetric gauge theories in 
$\mathcal{N} =1$ superspace have also been discussed \cite{pq}.

It is possible to define a  Polyakov  connection  on loop space which  measures the 
  change in phase  as one moves from one point in the loop space 
to a neighboring point. It is also possible to construct a  curvature tensor using this 
connection \cite{p2,p3}. This curvature is proportional to the  Bianchi 
identities and thus vanishes when the Bianchi identities are satisfied \cite{p4}. 
As in the presence of a  monopoles, Bianchi identities are not satisfied, so this 
curvature only gets a non-zero value when a monopole is present. Furthermore, it is possible to define 
a loop in the 
loop space which   covers a surface in spacetime. This loop in the loop space can be used  as a measure for the 
 non-abelian monopole charge. These results are know to hold for ordinary Yang-Mills theories. 
We shall derive them for the ABJ theory. We shall also generalize some of the previously known results. 
So, we shall obtain a curvature and connection in the space of loop of loops and use them for analysing 
topological defects in the loop space.

\section{Polyakov Loops }

As the 
ABJ theory is a Chern-Simons-Matter theory with  
the gauge group $U(N) \times U(M)$, so, we will denote the gauge 
fields corresponding to $U(N)$ by $A_\mu$ and the gauge fields 
corresponding to $U(M)$  by $A' _\mu$. These gauge fields are coupled to
complex scalar fields $C_I$ and their complex conjugates  $\bar C^I$, where $I = 1..4$ is an $SU(4)_R$ index.
They are also coupled to 
 fermions $\phi^a_I$ and $\bar \phi^I_a$, where $a = \pm$ is a spinor index.
 It may be noted that the matter fields $C_I, \bar\phi^I_a$ transforms under $(N, \bar M)$
 and the matter fields $\bar C^I, \phi^a_I$ transforms under $(\bar N, M)$ representations of the gauge group 
 $U(N) \times U(M)$. 
 We choose a notation such that $\bar C^I C_I$ and $\phi^a_I \bar\phi^I_a$ are in the adjoint representation 
 of $U(N)$ and $C_I \bar C^I $ and $\bar\phi_I^a \phi_a^I $ is in the adjoint representation of $U(M)$. 
 These fields for the ABJ theory transform under a superconformal transformations as follows
\begin{eqnarray}
\delta A_\mu&=&\frac{4\pi i}{k}\bar{\Theta}^{IJ\alpha}(\gamma_\mu)_\alpha^{\ \beta}\left(C_I\Psi_{J\beta}
+\frac{1}{2}\epsilon_{IJKL}\bar{\Psi}_\beta^K\bar{C}^L\right),
\nonumber \\
\delta  A'_\mu&=& \frac{4\pi i}{k}\bar{\Theta}^{IJ\alpha}(\gamma_\mu)_\alpha^{\ \beta}\left(\Psi_{J\beta}C_I
+\frac{1}{2}\epsilon_{IJKL}\bar{C}^L\bar{\Psi}_\beta^K\right),
\nonumber \\ 
\delta C_K&=&\bar{\Theta}^{IJ\alpha}\epsilon_{IJKL}\bar{\Psi}^L_\alpha,
\nonumber \\ 
\delta \bar{C}^K&=&2\bar{\Theta}^{KL\alpha}\Psi_{L\alpha},
\nonumber \\
\delta\Psi^\beta_K&=&-i\bar{\epsilon}^{IL\beta}\epsilon_{ILKJ}\bar{C}^J
-i\bar{\Theta}^{IJ\alpha}\epsilon_{IJKL}(\gamma^\mu)_\alpha^{\ \beta}D_\mu \bar{C}^{L} \nonumber \\ &&
+\frac{2\pi i}{k}\bar{\Theta}^{IJ\beta}\epsilon_{IJKL}(\bar{C}^LC_P\bar{C}^P-\bar{C}^PC_P\bar{C}^L)
\nonumber \\ &&+\frac{4\pi i}{k}\bar{\Theta}^{IJ\beta}\epsilon_{IJML}\bar{C}^MC_K\bar{C}^L, \nonumber\\
\delta\bar{\Psi}_\beta^K&=&-2i\bar{\Theta}^{KL\alpha}(\gamma^\mu)_{\alpha\beta}D_\mu C_{L}-2i\bar{\epsilon}^{KL}_\beta C_L\nonumber \\ &&
-\frac{4\pi i}{k}\bar{\Theta}^{KL}_\beta(C_L\bar{C}^MC_M-C_M\bar{C}^MC_L)\nonumber \\ &&
-\frac{8\pi i}{k}\bar{\Theta}^{IJ}_\beta C_I\bar{C}^KC_J.
\end{eqnarray}

 As we  want to study Polyakov loops for the ABJ theory,
we consider all the loops passing through some fixed point in spacetime,  
\begin{equation}
 C : \{ \xi^\mu (s): s = 0 \to 2\pi, \, \, \xi^\mu (0) = \xi^\mu (2\pi)\},  
\end{equation}
where $\xi^\mu (s) $ represents the spacetime coordinates of all points on the loop. We also define 
 $\dot{\xi}^\mu = {d \xi^\mu}/{ ds}$,
and $|\dot{\xi}|= \sqrt{\eta_{\mu\nu} \dot{\xi}^\mu  \dot{\xi}^\nu}$.
Even though, the gauge group of the ABJ theory is $U(N) \times U(M)$, we will embed it into a superconnection
 $\mathcal{A}$ belonging 
to $U(N|M)$ \cite{5}.
Scalar fields occur as bi-linears  because in three dimensions the dimension of scalar fields is $1/2$.
As the bi-linear combinations of scalar fields is in adjoint representation, they occur with the gauge 
fields in the diagonal blocks. 
Furthermore, as the dimensions of the fermions in three dimensions is $1$, they appear linearly. 
As the fermions  transform under bi-fundamental representation, they are placed off-diagonally. 
We define $M^I_J, {M'}^I_J, \eta^a_I, \bar\eta_a^I$ as the 
parameters in the theory which parameterize the local couplings. 
Even though $\eta^a_I, \bar\eta_a^I$ transform under a spinor representation of the Lorentz group, they are taken to 
be Grassmann even quantities. 
This is because by taking them to be Grassmann even quantities the off-diagonal entries become  Grassmann odd quantities. 
So, the superconnection for $U(N|M)$ can be written as, 
\begin{eqnarray}
 \mathcal{A} [\xi] = \left(
 \begin{array}{cc}
 \mathcal{A}_{11} [\xi] &   \mathcal{A}_{12} [\xi] \\  \mathcal{A}_{21} [\xi]
&  \mathcal{A}_{22} [\xi]  \\
 \end{array} \right), 
\end{eqnarray}
where 
\begin{eqnarray}
 \mathcal{A}_{11} [\xi] &=& A_\mu \dot{\xi}^\mu + \frac{2\pi}{k}|\dot{\xi}| M^I_J C_I \bar C^J,
 \nonumber \\
 \mathcal{A}_{12} [\xi] &=& \sqrt{\frac{2\pi}{k}}|\dot{\xi}|\eta^a_I \bar \phi_a^I, 
 \nonumber \\  
  \mathcal{A}_{21} [\xi] &=& \sqrt{\frac{2\pi}{k}}|\dot{\xi}| \phi^a_I \bar\eta_a^I,
  \nonumber \\
 \mathcal{A}_{22} [\xi] &=&  A'_\mu \dot{\xi}^\mu + \frac{2\pi}{k} |\dot{\xi}|{M'}^I_J  \bar C^J C_I.
\end{eqnarray}
So, we can write the field strength for this theory as $\mathcal{F}[\xi] 
=d \mathcal{A} [\xi] + \mathcal{A} [\xi]\wedge \mathcal{A} [\xi]$. The Bianchi identity can now be written as 
$(d + A [\xi]\wedge )\mathcal{F}[\xi] =0$.

In the dual string picture, 
the operators describing semi-classical string surfaces have  a local $U(1)\times SU(3)$  $R$-symmetry.
So, the $R$-symmetry  of the couplings can be described by a vector $n_I$ and its 
complex conjugate $\bar n_{I}$ \cite{abcd}. These specify the local embedding of 
of $SU(3)$ subgroup into $SU(4)$. They satisfy $n_{I}\bar n^{I}=1$.
 Now we have, $\eta_{I}^{\alpha} =n_{I} \eta^{\alpha},\,
 \bar\eta^{I}_{\alpha} =\bar n^{I}
 \bar\eta_{\alpha},\,    
M_{J}^{I} =p_{1} \delta^{I}_{ J}-2 p_{2}  n_{J} \bar n^{I },\,
 {M'}_{J}^{ I}=q_{1} \delta^{I}_{J}-2 q_{2}  n_{J}  \bar n^{I} $.
 The Eigenvalues of the $M_{J}^{I}$ and $ {M'}_{J}^{ I}$
 are controlled the  functions $p_{i}$ and $q_{i}$. 
The  condition that the supersymmetric variation of the superconnection vanishes is  
  too strong and it does not yield any solution for the  couplings.  
So, it is replaced by the requirement that the supersymmetry variation of the superconnection 
is equal to the covariant derivative generated from it.

The  spinor couplings are given by 
$\delta^{\beta}_{\alpha}=  (\eta^{\beta} \bar\eta_{\alpha}-\eta_{\alpha} \bar\eta^{\beta})/{2 i}$
and ${(\dot{x}^{\mu}\gamma_{\mu})_{\alpha}^{ \beta}}=\ell |\dot x|(\eta^{\beta} \bar\eta_{\alpha}+
\eta_{\alpha} \bar\eta^{\beta})/{2i}$. 
Furthermore, we have $M_{J}^{I} = {M'}_{J}^{ I} = \ell(\delta^{J}_{K}-2 n_{K}\bar n^{J})$. Here 
$\ell = \pm1$ and specifies  the eigenvalues of these matrices. 
Now $\epsilon_{IJKL} (\eta\bar{\Theta}^{IJ})\bar n^{K}=0$ and $n_{I}(\bar\eta\bar\Theta^{IJ})=0$ 
are the constraints on  
$\bar\Theta^{IJ}$. Apart from these constraints, it also satisfied 
$\bar{\Theta}^{IJ}(d/ds) {\bar{\eta}}^K\epsilon_{IJKL}=0$ and $\bar\Theta^{IJ}(d/ds) {\eta}_{I}=0$. 
These conditions are local and 
a conformal Killing spinor which satisfies these constraints has to be constructed for
obtaining a  supersymmetric  Polyakov loop.
If $\bar \theta^{IJ}$ and $\bar\epsilon^{IJ}$ are constant spinors, then we can write, 
$\bar\Theta^{IJ}=\bar\theta^{IJ}-( \gamma^\mu \xi_\mu)\bar\epsilon^{IJ}$.

Recall that the Polyakov 
loop \cite{p1} by the very definition is an element of the gauge group. 
Now the
Polyakov loop variables for the ABJ will be given by 
\begin{eqnarray}
 \phi [\xi] &=&  
\left(
 \begin{array}{cc}
 \phi _{11} [\xi ] &  \phi _{12} [\xi ] \\  \phi _{21} [\xi ]
&  \phi _{22} [\xi ]  \\
 \end{array} \right)
 \nonumber \\&=&  P_s \exp \int  ds
 \left(
 \begin{array}{cc}
 \mathcal{A}_{11} [\xi] &   \mathcal{A}_{12} [\xi] \\  \mathcal{A}_{21} [\xi]
&  \mathcal{A}_{22} [\xi]  \\
 \end{array} \right).
\end{eqnarray}
Here the ordering from right to left in $s$ 
is denoted by $P_s$.
It may be noted that $\phi[\xi]$ depends only on the loop $C$ in spacetime and not in 
the manner in which it is parametrized. 
If we introduce a new parameter say, $s' = f (s)$, it will only give a change in the variable of integration 
and not its value. 
So, at first sight it might appear  better to define the loops as equivalence
classes of the function $\xi(s)$, 
equivalent under reparametrization. But then it would be very
difficult to define differentiation 
and integration in this quotient space of equivalence classes, and
hence we will retain the original definition 
of the parametrized loops.

\section{ Connection and Curvature}
In this section we will construct a connection and a curvature for the
loop space. 
Strictly speaking, these do not have the exact geometric meanings of
the corresponding concepts in fiber bundles \cite{zois}, but the
formulae obtained below make sense in the context of loop space
variables and we shall continue to use these terms with this understanding.
We will first obtain a connection in the loop space and relate it to the field strength in spacetime. 
Then, we will construct a covariant derivative using this
connection. Finally, 
we will construct the curvature in the loop 
space from the commutator of these covariant derivatives. Now we first construct the connection in the loop 
space from $\phi [\xi]$, by taking its logarithmic derivative. As $\phi [\xi]$ is an element of the gauge group, 
its logarithmic derivative will be an element of the Lie algebra corresponding to that gauge group. 
So, we define the connection generated from $\phi [\xi]$ as follows, 
\begin{eqnarray}
 F_\mu[ \xi|s] &=&  
\left(
 \begin{array}{cc}
 F_\mu[ \xi|s] _{11}  &   F_\mu[ \xi|s] _{12}  \\   F_\mu[ \xi|s] _{21} 
&  F_\mu[ \xi|s] _{22}   \\
 \end{array} \right), 
\end{eqnarray}
where 
\begin{equation}
 F_\mu[ \xi|s] = i \phi^{-1} [\xi] \frac{\delta}{\delta \xi^\mu (s)}\phi [\xi]. 
\end{equation}
As $ F_\mu[ \xi|s]$ represents the change 
in $\phi[\xi]$ as one moves from one point in the loop space to its neighboring point, 
it can be regarded as a connection in parametrized loop space.
In calculations it is sometimes useful to define further 
$\phi [\xi(s_1,s_2)]$ as a 
parallel transport 
from a point $\xi (s_1)$ to a point  $\xi (s_1)$ along the curve 
$C$, 
\begin{eqnarray}
 \phi [\xi(s_1, s_2)] &=& 
\left(
 \begin{array}{cc}
 \phi _{11} [\xi (s_1, s_2)] &  \phi _{12} [\xi (s_1, s_2)] \\  \phi _{21} [\xi (s_1, s_2)]
&  \phi _{22} [\xi (s_1, s_2)]  \\
 \end{array} \right)
 \nonumber \\
\nonumber \\ &=& P_s  \exp \int_{s_1}^{s_2} ds 
\left(
 \begin{array}{cc}
 \mathcal{A}_{11} [\xi (s)] &   \mathcal{A}_{12} [\xi (s)] \\  \mathcal{A}_{21} [\xi (s)]
&  \mathcal{A}_{22} [\xi (s)]  \\
 \end{array} \right).
\end{eqnarray}

Now using $\phi [\xi (s_1, s_2)]$, we can move  from a fixed point
another point say, $s$, and then take a detour and travel backwards along the same path to the original point.
In doing this the phase factor generated in going from the original point to $s$, exactly 
cancels the phase factor generated in going back from $s$ to the original point. 
However, the phase factor while transporting around the infinitesimal circuit at $s$ does have a finite 
contribution that does not cancel. In fact, this contribution is proportional to the field strength 
$\mathcal{F}$. Thus,  $ F_\mu[\xi |s]$ is proportional to 
$ \phi^{-1} [\xi(s, 0)] \mathcal{F}[\xi (s)] \phi [\xi( s, 0)]
$. In fact, it is already know that in Yang-Mills theories the connection in loop space 
is proportional to the field strength in spacetime \cite{p2}.
We have observed here that this also hold for the superconnection of the ABJ theory.

In the loop space $  F_\mu[\xi]$ acts like a connection. The natural quantity to construct from this connection is the 
curvature of the loop space.  
Now we can define a covariant derivative in the loop space as follows, 
\begin{equation}
 \nabla_\mu [\xi(s)] =  \frac{\delta}{\delta \xi^\mu (s) } + i F_\mu [\xi|s].
\end{equation}
The curvature $ -i G_{\mu\nu}[\xi, s_1, s_2]$ 
of the loop space can be defined by taking a commutator of these two covariant derivatives, 
$
[\nabla_\mu [\xi(s_1)], \nabla_\nu [\xi(s_2)]]
$. Thus, we can write 
\begin{eqnarray}
 G_{\mu\nu}[\xi ( s_1, s_2)] &=& \frac{\delta}{\delta \xi^\mu (s_2) }F_\nu [\xi|s_1]
- \frac{\delta}{\delta \xi^\nu (s_1) }F_\mu [\xi|s_2] \nonumber \\&&
+i [F_\mu [\xi|s_1], F_\mu [\xi|s_2]].
\end{eqnarray}
 The  gauge transformations in loop space can be denoted by 
given by $u = \exp i \Lambda[\xi] $. The connection $F_\mu [\xi|s]$ transforms under 
these gauge transformations as $F_\mu[\xi|s] = i u \nabla_\mu [\xi(s)]u^{-1} $ 
and $G_{\mu\nu}[\xi( s_1, s_2 )]$ transforms under 
these gauge transformations as  $u G_{\mu\nu}[\xi ( s_1, s_2 )] u^{-1}$.

Now if we first 
travel from point say $s_1$ along a certain direction till a point say $s_2$. After that we travel along another direction 
at $s_1$, then we travel along the first direction and finally again travel along the direction we traveled from $s_2$, 
to get to $s_1$. In doing so we completed a full circuit and the total change in phase generated in the process 
is represented is proportional to 
$\phi^{-1} [\xi(s_1, 0)] \nabla^*\mathcal{F}[\xi(s_1)] \phi [\xi( s_1, 0 )]\delta (s_1-s_2)$
\cite{p4}. We have observed here that this also hold for the superconnection of the ABJ theory. 
Now this is also the value of by $-i G_{\mu\nu} [\xi (s_1, s_2)] \delta \xi^\mu (s_1) \xi^\nu (s_2)$. 
Hence, the curvature is proportional to $\phi^{-1} [\xi(s_1, 0)] \nabla^*\mathcal{F}[\xi (s_1)] \phi [\xi( s_1, 0)]
\delta (s_1-s_2)$.
Thus, if the Bianchi 
identity is satisfied $ \nabla^*\mathcal{F}[\xi (s_1)] =0$, 
this curvature vanishes $G_{\mu\nu} [\xi(s_1, s_2)] =0$. 
However, in presence of a monopole, Bianchi identity is not satisfied and thus this curvature does not vanish. 

It may be noted that $ G_{\mu\nu}[\xi( s_1, s_2)]$ satisfies 
a functional Bianchi identity even in presence of a monopole. 
To derive this functional Bianchi identity, we first  define
$\nabla_\mu [\xi(s_1)]^* [\nabla_\nu [\xi(s_2)], \nabla_\rho [\xi(s_3)]]$ as follows, 
\begin{eqnarray}
&& \nabla_\mu [\xi(s_1)]^* [\nabla_\nu [\xi(s_2)], \nabla_\rho [\xi(s_3)]]\nonumber \\
 &=& \nabla_\rho [\xi(s_3)][\nabla_\mu [\xi(s_1)], \nabla_\nu [\xi(s_2)]]  \nonumber \\ && + 
   \nabla_\nu [\xi(s_2)] [\nabla_\rho [\xi(s_3)], \nabla_\mu [\xi(s_1)]]\nonumber \\ &&
    + \nabla_\mu [\xi(s_1)] [\nabla_\nu [\xi(s_2)], \nabla_\rho [\xi(s_3)]]. 
    \end{eqnarray}
Now, expanding this expression for $\nabla_\mu [\xi(s_1)]^* [\nabla_\nu [\xi(s_2)], \nabla_\rho [\xi(s_3)]]$, 
we get 
\begin{eqnarray}
&&\nabla_\mu [\xi(s_1)]^* [\nabla_\nu [\xi(s_2)], \nabla_\rho [\xi(s_3)]]\nonumber \\
&=&
   \left( \frac{\delta}{\delta \xi^\rho (s_3) } + i F_\rho [\xi|s_3]\right)
\left( \frac{\delta}{\delta \xi^\mu (s_1) } + i F_\mu [\xi|s_1]\right) 
\left(
\frac{\delta}{\delta \xi^\nu (s_2) } + i F_\nu [\xi|s_2]\right) \nonumber \\ && 
-
\left( \frac{\delta}{\delta \xi^\rho (s_3) } + i F_\rho [\xi|s_3]\right)
\left(\frac{\delta}{\delta \xi^\nu (s_2) } + i F_\nu [\xi|s_2]\right) 
\left( \frac{\delta}{\delta \xi^\mu (s_1) } + i F_\mu [\xi|s_1]\right)
\nonumber \\ && + 
  \left( \frac{\delta}{\delta \xi^\nu (s_2) } + i F_\nu [\xi|s_2]\right)\left(
  \frac{\delta}{\delta \xi^\rho (s_3) } + i F_\rho [\xi|s_3]\right)
  \left( 
   \frac{\delta}{\delta \xi^\mu (s_1) } + i F_\mu [\xi|s_1]\right)
   \nonumber \\ &&
   -
     \left( \frac{\delta}{\delta \xi^\nu (s_2) } + i F_\nu [\xi|s_2]\right)\left( 
   \frac{\delta}{\delta \xi^\mu (s_1) } + i F_\mu [\xi|s_1]\right)
\left(
  \frac{\delta}{\delta \xi^\rho (s_3) } + i F_\rho [\xi|s_3]\right)
   \nonumber \\ &&   
   + \left(
    \frac{\delta}{\delta \xi^\mu (s_1) } + i F_\mu [\xi|s_1]\right)
    \left( \frac{\delta}{\delta \xi^\nu (s_2) } + i F_\nu [\xi|s_2]\right) 
    \left(\frac{\delta}{\delta \xi^\rho (s_3) } + i F_\rho [\xi|s_3]\right) 
   \nonumber \\ &&-
   \left(
    \frac{\delta}{\delta \xi^\mu (s_1) } + i F_\mu [\xi|s_1]\right)
    \left(\frac{\delta}{\delta \xi^\rho (s_3) } + i F_\rho [\xi|s_3]\right) 
       \left( \frac{\delta}{\delta \xi^\nu (s_2) } + i F_\nu [\xi|s_2]\right) 
   \nonumber \\ &=& 0, 
\end{eqnarray}
where 
\begin{eqnarray} &&  \frac{\delta}{\delta \xi^\mu (s_1) } + i F_\mu [\xi|s_1]
\nonumber \\&=&
  \begin{pmatrix}  \frac{\delta}{\delta \xi^\mu (s_1) }+ i F_\mu[ \xi|s_1] _{11}  &  i F_\mu[ \xi|s_1] _{12} 
  \\  i  F_\mu[ \xi|s_1] _{21}  &   \frac{\delta}{\delta \xi^\mu (s_1) }+ i F_\mu[ \xi|s_1] _{22} \end{pmatrix},
  \\
  &&  \frac{\delta}{\delta \xi^\nu (s_2) } + i F_\nu [\xi|s_2]
\nonumber \\&=&
  \begin{pmatrix}  \frac{\delta}{\delta \xi^\nu (s_2) }+ i F_\nu[ \xi|s_2] _{11}  &  i F_\nu[ \xi|s_2] _{12} 
  \\  i  F_\nu[ \xi|s_2] _{21}  &   \frac{\delta}{\delta \xi^\nu (s_2) }+ i F_\nu[ \xi|s_2] _{22} \end{pmatrix},
  \\ 
&& \frac{\delta}{\delta \xi^\rho (s_3) } + i F_\rho [\xi|s_3]
\nonumber \\&=&
  \begin{pmatrix}  \frac{\delta}{\delta \xi^\rho (s_3) }+ i F_\rho[ \xi|s_3] _{11}  &  i F_\rho[ \xi|s_3] _{12} 
  \\  i  F_\rho[ \xi|s_3] _{21}  &   \frac{\delta}{\delta \xi^\rho (s_3) }+ i F_\rho[ \xi|s_3] _{22} \end{pmatrix}.
\end{eqnarray}
Thus, we get,    $\nabla_\mu [\xi(s_1)]^* [\nabla_\nu [\xi(s_2)], \nabla_\rho [\xi(s_3)]] =0$. 

However, as the curvature of the loop space is generated by the commutator of the functional 
covariant derivatives, we observe that 
the functional Bianchi identity is satisfied for the loop space. 
So, we can write 
 $ \nabla_\rho [\xi(s_3)]G_{\mu\nu}[\xi( s_1, s_2)]
+\nabla_\mu [\xi(s_1)]G_{\nu\rho}[\xi ( s_2, s_3)] +\nabla_\nu [\xi(s_2)]G_{\rho\mu}[\xi ( s_3, s_1)]
=0 $. 
We emphasize again that in the presence of a monopole, the space-time
Bianchi identity is not satisfied, and this translates into the
non-vanishing of the loop space curvature $G_{\mu\nu} [\xi(s_1,
s_2)]$.
On the other hand, the loop space curvature itself does satisfy the Bianchi
identity even in the presence of a monopole.

\section{Loop of Loops}

In the previous section we analysed Polyakov loops for the ABJ theory.
It may be noted that Polyakov loops have been generalized to loop of loops for Yang-Mills theories \cite{p4}. 
Here we will apply this formalism of loop of loops to the ABJ theory. We will also extend this formalism to 
include the concept of a connection and curvature for loop of loops.  
A loop in the loop space can be defined using $ F_\mu[ \xi|s]$
as the connection. Now we can parameterize a loop in the loop space as follows 
\cite{p4},
\begin{equation}
  \Sigma : \{ \xi^\mu (t:s), \, s = 0 \to 2 \pi, \, t = 0 \to 2 \pi\},
\end{equation}
where    $
\xi^\mu (t:0) =  \xi^\mu (t:2\pi), $ and $  t  = 0 \to 2 \pi  $.
At each value of $t$, a closed loop $C(t)$ is traced in the spacetime passing through a fixed point. 
Thus, for $t=0$ and $t = 2\pi$ it shrinks to this fixed point and as $t$ varies from $0 \to 2\pi$, 
$C(t)$ traces out a closed loop in the loop space. This loop starts and ends at the fixed point to which 
$C(t)$ shrinks for $t=0$ and $t = 2\pi$. Thus, we can define a loop variable for this space as, 
\begin{eqnarray}
 \Theta  [\xi] &=&  
\left(
 \begin{array}{cc}
 \Theta  _{11} [\xi ] & \Theta  _{12} [\xi ] \\  \Theta  _{21} [\xi ]
& \Theta  _{22} [\xi ]  \\
 \end{array} \right) \nonumber \\
 &=&   P_t   \exp  i \int^{2\pi}_{0} dt \int^{2\pi}_0 ds 
\left(
 \begin{array}{cc}
 F^\mu[ \xi|t:s] _{11}  &   F^\mu[ \xi|t:s] _{12}  \\   F^\mu[ \xi|t:s] _{21} 
&  F^\mu[ \xi|t:s] _{22}   \\
 \end{array} \right)  \nonumber \\ && \times \frac{\partial \xi_\mu (t:s)}{ \partial t}, 
\end{eqnarray}
where $P_t$ denotes ordering in $t$ increasing from right to left and the derivative is taken from below. 
The connection in the loop space, $F^\mu[\xi|s]$,  plays the role
of the space-time gauge field ${\cal A}[\xi]$ in the loop space, so that
this definition is the analogue in loop space of (5).
However, the connection in the loop space is  infinite dimensional and so apart from the sum  
over $\mu$, we have to also integrate over $s$. 
In ordinary spacetime this parametrized loop in loop space is represented by a 
 two dimensional surface which enclosing a three dimensional 
volume.

In analogy with the previous case we can define a connection in this space using $\Theta [\xi]$. 
In fact, we will define the connection in this space to be the logarithmic derivative of $\Theta [\xi]$. 
So, we write 
\begin{eqnarray}
 B_\mu[ \xi|t: s]&=&  
\left(
 \begin{array}{cc}
 B_\mu[ \xi|t: s] _{11}  &   B_\mu[ \xi|t: s] _{12}  \\   B_\mu[ \xi|t: s] _{21} 
&  B_\mu[ \xi|t: s] _{22}   \\
 \end{array} \right), 
\end{eqnarray}
where 
\begin{equation}
 B_\mu[ \xi|t: s] = i \Theta^{-1} [\xi] \frac{\delta}{\delta \xi^\mu (t:s)}\Theta [\xi].
\end{equation}
Geometrically, $ B_\mu[ \xi|t : s]$ can be regarded as a connection in the space of loop of loops, 
as it represents the change 
in $\Theta[\xi]$ as one moves from one point in this  space to its neighboring point.
Now we can define a quantity which will act as parallel transport in this space 
\begin{eqnarray}
 \Theta [ \xi(t_1, t_2) ] &=&   
\left(
 \begin{array}{cc}
 \Theta  _{11} [\xi (t_1, t_2) ] & \Theta  _{12} [\xi (t_1, t_2) ] \\  \Theta  _{21} [\xi (t_1, t_2) ]
& \Theta  _{22} [\xi (t_1, t_2) ]  \\
 \end{array} \right) \nonumber \\ &=& P_t   \exp  i \int^{t_2}_{t_1} dt \int^{2\pi}_0 ds 
\left(
 \begin{array}{cc}
 F^\mu[ \xi|t:s] _{11}  &   F^\mu[ \xi|t:s] _{12}  \\   F^\mu[ \xi|t:s] _{21} 
&  F^\mu[ \xi|t:s] _{22}   \\
 \end{array} \right)  \nonumber \\ && \times \frac{\partial \xi_\mu (t:s)}{ \partial t}.
\end{eqnarray}
Now using $\Theta[\xi  (t_1, t_2)]$, we can move  from a fixed point
another point and then take a detour and travel backwards along the same path to then original path.
In doing this the phase factor for the generated in going from the original point to final point, exactly 
cancels the phase factor generated in going back from the final point to the original point. 
However, the phase factor while transporting around the infinitesimal circuit at the final point does have a finite 
contribution that does not cancel. This contribution is proportional to the curvature of the loop space.
In fact, by repeating the previous calculations, we observe that 
\begin{eqnarray}
 B_\mu [\xi(t_1: s_1)] &=&
 \int ds_2 \Theta^{-1}[\xi  (t_1, 0)]G_{\mu\nu}[\xi ( t_1: s_1, s_2 )] \Theta[\xi  (0, t_1)] 
 \nonumber \\  && \times \frac{\partial\xi^\nu(t_1:s_2)}{\partial t_1}.
\end{eqnarray}
So, 
$B^\mu [\xi(t_1: s_1)]$ is proportional to the curvature of the loop space. 
Now for $s_1 \neq s_2$, $G_{\mu\nu}[\xi ( t_1: s_1, s_2 )]$ corresponds to 
a parameterized surface enclosing no volume and $B^\mu [\xi(t_1: s_1)]$  in this case 
is zero. The same value is obtained for $s_1 = s_2$, if it the volume $\Sigma$ encloses does not contain a monopole. 

As in the space of loop of loops $B^\mu [\xi(t: s)]$ acts like a connection, we can construct a  
 covariant derivative using it, 
\begin{equation}
 \bar \nabla_\mu [\xi(t: s)] =  \frac{\delta}{\delta \xi^\mu (t: s) } + i B_\mu [\xi|t:s].
\end{equation}
We can now define a  curvature $ -i E_{\mu\nu}[\xi( t_1, t_2 : s_1, s_2)]$ 
of this space as follows 
$
[\bar \nabla_\mu [\xi(t_1: s_1)], \bar \nabla_\nu [\xi(t_2: s_2)]]
$. Thus, we can write 
\begin{eqnarray}
 E_{\mu\nu}[\xi ( t_1, t_2: s_1, s_2)] &=& \frac{\delta}{\delta \xi^\mu (t_2: s_2) }B_\nu [\xi|t_1: s_1]
- \frac{\delta}{\delta \xi^\nu (t_1: s_1) }B_\mu [\xi|t_2: s_2] \nonumber \\&&
+i [B_\mu [\xi|t_1:s_1], B_\mu [\xi|t_2: s_2]].
\end{eqnarray}
 The  gauge transformations in loop space can be denoted by 
given by $v = \exp i \Lambda[\xi] $. The connection $B_\mu [\xi|t: s]$ transforms under 
these gauge transformations as $B_\mu[\xi|t: s] = i v \bar \nabla_\mu [\xi(t:s)]v^{-1} $ 
and $E_{\mu\nu}[\xi( t_1, t_2: s_1, s_2 )]$ transforms under 
these gauge transformations as  $v E_{\mu\nu}[\xi (t_1, t_2: s_1, s_2 )] v^{-1}$.

It may be noted that $ E_{\mu\nu}[\xi(t_1, t_2:  s_1, s_2)]$ again  satisfies 
a functional Bianchi identity. 
Now we  define
$\bar \nabla _\mu [\xi(t_1: s_1)]^* [\bar \nabla _\nu [\xi(t_2: s_2)], \bar \nabla _\rho [\xi(t_3, s_3)]]$ as follows, 
\begin{eqnarray}&& \bar \nabla _\mu [\xi(t_1: s_1)]^* [\bar \nabla _\nu [\xi(t_2: s_2)], \bar \nabla _\rho [\xi(t_3: s_3)]]\nonumber \\
&=&  \bar \nabla _\rho [\xi(t_3: s_3)][\bar \nabla _\mu [\xi(t_1: s_1)], \bar \nabla _\nu [\xi(t_2: s_2)]]\nonumber \\ &&  + 
   \bar \nabla _\nu [\xi(t_2:s_2)] [\bar \nabla _\rho [\xi(t_3: s_3)], \bar \nabla _\mu [\xi(t_1: s_1)]]\nonumber \\ &&
    + \bar \nabla _\mu [\xi(t_1: s_1)] [\bar \nabla _\nu [\xi(t_2: s_2)], \bar \nabla _\rho [\xi(t_3:s_3)]]. 
    \end{eqnarray}
In order to prove the functional Bianchi identity for $ E_{\mu\nu}[\xi(t_1, t_2:  s_1, s_2)]$, we expand 
$\bar \nabla _\mu [\xi(t_1: s_1)]^* [\bar \nabla _\nu [\xi(t_2:s_2)], \bar \nabla _\rho [\xi(t_3: s_3)]]$, 
as follows
\begin{eqnarray}
&&\bar \nabla _\mu [\xi(t_1: s_1)]^* [\bar \nabla _\nu [\xi(t_2:s_2)], \bar \nabla _\rho [\xi(t_3: s_3)]]\nonumber \\
&=&
   \left( \frac{\delta}{\delta \xi^\rho (t_3: s_3) } + i B _\rho [\xi|t_3: s_3]\right)
\left( \frac{\delta}{\delta \xi^\mu (t_1: s_1) } + i B _\mu [\xi|t_1: s_1]\right)
\nonumber \\ &&\times 
\left(
\frac{\delta}{\delta \xi^\nu (t_2: s_2) } + i B _\nu [\xi|t_2: s_2]\right) \nonumber \\ && 
-
\left( \frac{\delta}{\delta \xi^\rho (t_3: s_3) } + i B _\rho [\xi|t_3: s_3]\right)
\left(\frac{\delta}{\delta \xi^\nu (t_2: s_2) } + i B _\nu [\xi|t_2: s_2]\right) \nonumber \\ &&\times 
\left( \frac{\delta}{\delta \xi^\mu (t_1: s_1) } + i B _\mu [\xi|t_1: s_1]\right)
\nonumber \\ && + 
  \left( \frac{\delta}{\delta \xi^\nu (t_2: s_2) } + i B _\nu [\xi|t_2: s_2]\right)\left(
  \frac{\delta}{\delta \xi^\rho (t_3: s_3) } + i B _\rho [\xi|t_3: s_3]\right)
  \nonumber \\ &&\times 
  \left( 
   \frac{\delta}{\delta \xi^\mu (t_1: s_1) } + i B _\mu [\xi|t_1: s_1]\right)
   \nonumber \\ &&
   -
     \left( \frac{\delta}{\delta \xi^\nu (t_2: s_2) } + i B _\nu [\xi|t_2: s_2]\right)\left( 
   \frac{\delta}{\delta \xi^\mu (t_1: s_1) } + i B _\mu [\xi|t_1: s_1]\right)
   \nonumber \\ &&\times \left(
  \frac{\delta}{\delta \xi^\rho (t_3: s_3) } + i B _\rho [\xi|t_3: s_3]\right)
   \nonumber \\ &&   
   + \left(
    \frac{\delta}{\delta \xi^\mu (t_1: s_1) } + i B _\mu [\xi|t_1: s_1]\right)
    \left( \frac{\delta}{\delta \xi^\nu (t_2: s_2) } + i B _\nu [\xi|t_2: s_2]\right) \nonumber \\ &&\times 
    \left(\frac{\delta}{\delta \xi^\rho (t_3: s_3) } + i B _\rho [\xi|t_3: s_3]\right) 
   \nonumber \\ &&-
   \left(
    \frac{\delta}{\delta \xi^\mu (t_1: s_1) } + i B _\mu [\xi|t_1: s_1]\right)
    \left(\frac{\delta}{\delta \xi^\rho (t_3: s_3) } + i B _\rho [\xi|t_3: s_3]\right)\nonumber \\ &&\times  
       \left( \frac{\delta}{\delta \xi^\nu (t_2: s_2) } + i B _\nu [\xi|t_2: s_2]\right) 
   \nonumber \\ &=& 0,  
\end{eqnarray}
where  
\begin{eqnarray} &&  \frac{\delta}{\delta \xi^\mu (t_1 : s_1) } + i B_\mu [\xi|t_1 : s_1]
\nonumber \\&=&
  \begin{pmatrix}  \frac{\delta}{\delta \xi^\mu (t_1 : s_1) }+ i B_\mu[ \xi|t_1 : s_1] _{11}  &  i B_\mu[ \xi|t_1 : s_1] _{12} 
  \\  i  B_\mu[ \xi|t_1 : s_1] _{21}  &   \frac{\delta}{\delta \xi^\mu (t_1 : s_1) }+ i B_\mu[ \xi|t_1 : s_1] _{22} \end{pmatrix},
  \\
  &&  \frac{\delta}{\delta \xi^\nu (t_2 : s_2) } + i B_\nu [\xi|t_2 : s_2]
\nonumber \\&=&
  \begin{pmatrix}  \frac{\delta}{\delta \xi^\nu (t_2 : s_2) }+ i B_\nu[ \xi|t_2 : s_2] _{11}  &  i B_\nu[ \xi|t_2 : s_2] _{12} 
  \\  i  B_\nu[ \xi|t_2 : s_2] _{21}  &   \frac{\delta}{\delta \xi^\nu (t_2 : s_2) }+ i B_\nu[ \xi|t_2 : s_2] _{22} \end{pmatrix},
  \\ 
&& \frac{\delta}{\delta \xi^\rho (t_3 : s_3 ) } + i B_\rho [\xi|t_3 : s_3 ]
\nonumber \\&=&
  \begin{pmatrix}  \frac{\delta}{\delta \xi^\rho (t_3 : s_3 ) }+ i B_\rho[ \xi|t_3 : s_3 ] _{11}  &  i B_\rho[ \xi|t_3 : s_3 ] _{12} 
  \\  i  B_\rho[ \xi|t_3 : s_3 ] _{21}  &   \frac{\delta}{\delta \xi^\rho (t_3 : s_3 ) }+ i B_\rho[ \xi|t_3 : s_3 ] _{22} \end{pmatrix}.
\end{eqnarray}
Thus, we get,    $\bar \nabla _\mu [\xi(t_1: s_1)]^* [\bar \nabla _\nu [\xi(t_2: s_2)],
\bar \nabla _\rho [\xi(t_3: s_3)]] =0$. 

However, as the curvature of the loop space is generated by the commutator of the functional 
covariant derivatives, we observe that 
the functional Bianchi identity is satisfied for this space. 
So, we can write 
 $ \bar \nabla _\rho [\xi(t_3: s_3)]E_{\mu\nu}[\xi( t_1, t_2: s_1, s_2)]
+\bar \nabla _\mu [\xi(t_1: s_1)]E_{\nu\rho}[\xi ( t_2, t_3: s_2, s_3)]
+\bar \nabla _\nu [\xi(t_2: s_2)]E_{\rho\mu}[\xi ( t_3, t_1: s_3,  s_1)]
=0
$.

 \section{Topological Defects}
 The both the ABJM theory and the ABJ theory have $\mathcal{N} =6$ supersymmetry. 
 However, it is expected that for ABJM theory for the Chern-Simons levels, $k =1, 2$, this supersymmetry will get 
 enhanced to $\mathcal{N} =8$ supersymmetry \cite{mp, mp1, mp2, mp0}. In this supersymmetric enhancement an important role 
 is played by the monopole operators. Thus, it is important to understand the role of monopoles in the ABJM theory. 
 In fact, in this section we will analyse the monopoles in the ABJ theory. We will also 
 study a topological defect in the loop space.  This defect in loop space is similar to a monopole in spacetime.

 So, now we will analyse monopoles in the ABJ theory. To do that, we first note that 
 whenever $\mathcal{F}$ is derivable from the superconnection $\mathcal{A}$, Bianchi identities for $\mathcal{F}$
 will be satisfied, $\nabla ^* \mathcal{F}=0$. As the curvature of the loop space is proportional to 
 the Bianchi identities, it will vanish whenever $\mathcal{F}$ is derivable from the superconnection $\mathcal{A}$. 
 However, at a point where the loop intersects the world-line of a monopole, 
 $\mathcal{F}$ will not be derivable from the superconnection $\mathcal{A}$ and 
 the Bianchi identities will not hold. Thus, the above argument will not hold and the curvature can get a  non-zero 
 value. In other words if $ G_{AB}[\xi( s_1, s_2)] \neq 0$ then $\nabla ^* \mathcal{F}\neq 0$ and the loop  will be 
intersecting word-lines of a monopole.

As $\Theta $ measure the total change in the loop as $t = 0 \to 2 \pi$, so, if a monopole is present 
it will not wind fully around the gauge group. However, in absence of a monopole, it will wind fully around the 
gauge group. Thus, we can write \cite{p4} $\Theta = \zeta I$, where $\zeta$ is the monopole charge of the ABJ theory 
enclosed by the surface $\Sigma$. Thus, the monopole charge corresponds to the loop in the loop space for the ABJ theory.
If the loop passes through a monopole, then at the value of $s_1$, where the loop $\xi(s_1)$ intersects  
the monopole 
world-line $Y(s_3)$, the curvature will not vanish. In fact, it will be given by \cite{p4}
\begin{eqnarray}
 G_{\mu\nu}[\xi ( s_1, s_2 )] &=& - \pi \int ds_3 \kappa [\xi|s]\epsilon_{\mu\nu \rho \tau} \frac{d \xi^\rho (s_1)}{ds_1} 
 \frac{d\xi^\tau (s_3)}{ds_3} \nonumber \\ && \times \delta^3 (\xi (s_1) - Y(s_3)) \delta(s_1-s_2).
\end{eqnarray}
Here $\kappa [\xi |s]$ satisfies $\exp i \pi \kappa = \zeta$, where, $\zeta$ is the charge carried by 
the monopole moving along the world-line $Y(s_3)$.

We have observed 
that even when monopoles are present a functional Bianchi identity for the curvature in the loop 
space is satisfied. Furthermore, for the space of loop of loops, let us  first 
travel from point say $t_1$ along a certain direction till a point say $t_2$. After that we travel along another direction 
at $t_1$, then we travel along the first direction and finally again travel along the direction we traveled from $t_2$, 
to get to $t_1$. In doing so we completed a full circuit and the total change in phase generated in the process 
is represented is proportional to the curvature $-i E_{\mu\nu} [\xi (t_1, t_2: s_1, s_2)]
\delta \xi^\mu (t_1: s_1) \xi^\nu (t_2: s_2)$. Now we calculate the quantity 
given by $\Theta^{-1}[ \xi_2] \Theta [\xi_3] - \Theta [\xi] \Theta [\xi_1] $, 
where $\xi_1 ^{\mu}[t:s] = \xi^{\mu}[t:s]  + \delta \xi^{\mu}[t:s], 
\xi_2^{\mu}[t:s] = \xi^{\mu}[t:s] + \delta' \xi^{\mu}[t:s], \xi_3^{\mu}[t:s] = 
\xi_1^{\mu}[t:s] + \delta'\xi^{\mu}[t:s]$. 
Now we can write 
\begin{eqnarray}
 \Theta [\xi_1] &=& \Theta[\xi] - ig \int dt \int ds_1 ds_2 \Theta[ \xi (2\pi, t)] G_{\mu\nu } [\xi( t: s_1, s_2)]  
 \nonumber \\&& \times \frac{\partial \xi^\nu (t: s_2)}{\partial t} \delta \xi^\mu (t,s_1) \Theta[\xi(t, 0)]. 
\end{eqnarray}
We also have 
\begin{eqnarray}
 \Theta [\xi_2] &=& \Theta[\xi] - ig \int dt \int ds_1 ds_2 \Theta[ \xi (2\pi, t)] G_{\mu\nu } [\xi( t: s_1, s_2)]  
 \nonumber \\&& \times \frac{\partial \xi^\nu (t: s_2)}{\partial t} \delta' \xi^\mu (t,s_1) \Theta[\xi(t, 0)]. 
\end{eqnarray}
Finally, we have 
\begin{eqnarray}
 \Theta [\xi_3] &=& \Theta[\xi_1] - ig \int dt \int ds_1 ds_2 \Theta[ \xi_1 (2\pi, t)] G_{\mu\nu } [\xi_1( t: s_1, s_2)]  
 \nonumber \\&& \times \frac{\partial \xi^\nu_1 (t: s_2)}{\partial t} \delta' \xi^\mu_1 (t,s_1) \Theta[\xi_1(t, 0)]. 
\end{eqnarray}
We also note that 
\begin{eqnarray}
 G_{\mu\nu}[\xi_1(t: s_1, s_2)] &=&  
  \int ds_3 \frac{\delta }{\delta \xi^{\rho} (t: s_3)} G_{\mu\nu}[\xi(t: s_1, s_2)] \delta \xi^{\rho} (t: s_3)
 \nonumber \\ && +  G_{\mu\nu}[\xi(t: s_1, s_2)].
\end{eqnarray}
Now  by collection all the terms for $E_{\mu\nu}[\xi( t_1, t_2: s_1, s_2)]$ is given by 
\begin{eqnarray}
 E_{\mu\nu}[\xi( t_1, t_2: s_1, s_2)] &=&\int ds_3 \Theta^{-1}[\xi(t_1, 0) ] 
  [ \nabla_\rho [\xi(t_1: s_3)]G_{\mu\nu}[\xi( t_1: s_1, s_2)]
 \nonumber \\ && +\nabla_\mu [\xi(t_1: s_1)]G_{\nu\rho}[\xi ( t_1: s_2, s_3)]  \nonumber \\ && +\nabla_\nu [\xi(t_1: s_2)]
G_{\rho\mu}[\xi ( t_1: s_3, s_1)]] 
 \nonumber \\ 
 && \times \Theta[\xi(t_1, 0) ]  \frac{\partial \xi^{\rho}(t_1: s_3)}{\partial t_1}\delta(t_1 -t_2).
\end{eqnarray}
So, the curvature of the space of loop of loops is proportional to the functional Bianchi identity, 
$\nabla_\rho [\xi(t_1: s_3)]G_{\mu\nu}[\xi( t_1: s_1, s_2)]
+\nabla_\mu [\xi(t_1: s_1)]G_{\nu\rho}[\xi ( t_1: s_2, s_3)] +\nabla_\nu [\xi(t_1: s_2)]
G_{\rho\mu}[\xi ( t_1: s_3, s_1)]$.

If this functional Bianchi identity is satisfied, then this curvature
vanishes. However, one can envisage a singularity in loop space which
is similar to the singularity in space-time giving a monopole.
 If such a monopole like defect occurs, such that the functional Bianchi identity is not satisfied, 
then the curvature for loop of loops will not be zero. It would be interesting to see what would be the possible 
implications of such a topological defect. 
It may be noted that even in presence of such a defect, 
the functional Bianchi identity for $E_{\mu\nu}[\xi( t_1, t_2: s_1, s_2)]$ hold, 
$ \bar\nabla_\rho [\xi(t_3: s_3)]E_{\mu\nu}[\xi( t_1, t_2: s_1, s_2)]
+\bar\nabla_\mu [\xi(t_1:s_1)]E_{\nu\rho}[\xi ( t_2, t_3: s_2, s_3)]
+\bar \nabla_\nu [\xi(t_2: s_2)]E_{\rho\mu}[\xi ( t_3, t_1: s_3, s_1)] =0
$.

\section{Conclusion}

In this paper we analysed Polyakov loops for the ABJ theory. This was done by first constructing the loop variables 
in terms of a superconnection. The fermions coupled to the loop in the bi-fundamental representation. 
Then a connection for this loop was constructed and a curvature tensor from this connection was also constructed. 
This curvature tensor was found to be proportional to the Bianchi identities and thus vanished when they were satisfied. 
As the Bianchi identities are not satisfied in presence of a monopole, so this curvature tensor  
has a non-vanishing value  in the presence of a monopole. 
A space of loop of loop  was also construed and was used as a measure for the monopole charge of the ABJ theory. 
We also constructed the curvature and connection for this space of loop of loops. 
Furthermore, certain topological defects in the loops space were analysed using the curvature of 
the space of loop of loops. 
It would be interesting to investigate further the physical implications of topological defect in the loop space. 
If a loop in the loop space is really the quantity which captures real  physics, then it seems likely that 
a topological defect in this space of loop of loops can have real physical meaning. Furthermore, it 
would be interesting to see how far can we go with such constructions. We have already constructed the 
functional Bianchi identity for the space of loop of loops. The next natural question to analyse 
is the existence of topological defects in this space which violates this functional  Bianchi identity. 
The we can perform a similar analyse for those defect too.

If we consider the ABJM  theory and give a vacuum expectation value to one of the scalar fields, then we 
arrive at the action for multiple D2-branes \cite{d2,sxsw, xswd, d21}. In this mechanism 
  the gauge group   $U(N) \times U(N)$ is broken down 
  to its diagonal subgroup. The theory thus obtained is  the Yang-Mills theory coupled to matter fields. 
It would be interesting to analyse  the Polyakov loops for the 
ABJM theory, after a vacuum expectation value  is given to one of the scalar fields. 
It would be expected that the Polyakov loops for the ABJM theory in this case will 
reduce to the Polyakov loop for the D2-branes. Furthermore, it will interesting to analyse what 
happens to monopole charge of the ABJM theory, after a vacuum expectation value  is given to one of the scalar fields. 
The Wilsons loops in generic tensor representation for IIB string theory are dual to D3-branes \cite{dual}. 

In fact, they are also dual to D5-branes \cite{2dual}. It is thus expected that string like objects in 
$AdS_3 \times CP^3$ will be
dual to Wilsons loops in different representation of  $SU(N|M)$ \cite{dual2}. It would be interesting to 
construct such operators for the ABJM theory and use them for perturbative calculations. 
It will also be interesting to analyse what happens to this duality when the Wilsons loops are replaced by Polyakov loops.

A system of M2-branes ending on other objects in M-theory has been studied by analysing the ABJM theory 
and the BLG theory 
on a manifold with boundaries \cite{mf, mf2, mf1}.
  Boundary conditions for the 
M2-branes ending  on M5-branes, M9-branes and gravitational waves have  been studied \cite{BCIntMem}.
Furthermore, a background flux can exist  in M-theory.
Boundary
conditions for M2-branes in the presence of a background flux  have also  been discussed \cite{ChuSehmbi}.
It is possible to learn about the physics of M5-branes by studding a system of M2-branes ending on them. 
A novel quantum geometry on M5-branes has been studied by analysing a system of M2-branes ending on 
a  M5-brane with constant  $C$-field \cite{d12}. The BLG theory was used to study this novel geometry. 
In fact, the BLG theory with Nambu-Poisson $3$-bracket has been identified with the   M5-brane action 
 with a large worldvolume C-field \cite{M5BLG}.

 A non-commutative string theory on the M5-brane worldvolume
has been derived using the action of a single M2-brane 
\cite{NCS1, NCS2, NCS3}.  It would be interesting to analyse the boundary effects for fractional M2-branes 
using ABJ theory. It would also be interesting to analyse topological defects in this theory using 
Polyakov loops.


\begin{thebibliography}{99}
\bibitem{1b}A. Gustavsson, JHEP. 0804, 083 (2008)
 \bibitem{2b}J. Bagger and N. Lambert, JHEP. 0802, 105 (2008)
\bibitem{3b}J. Bagger and N. Lambert, Phys. Rev. D77, 065008 (2008)
 \bibitem{4b}M. A. Bandres, A. E. Lipstein and J. H. Schwarz, JHEP. 0809, 027 (2008)
\bibitem{5b}E. Antonyan and  A. A. Tseytlin, Phys. Rev. D79, 046002 (2009)
\bibitem{abjm}O. Aharony, O. Bergman, D. L. Jafferis and J. Maldacena,
 JHEP. 0810, 091 (2008)
 \bibitem{ab}
 M. A. Bandres, A. E. Lipstein and J. H. Schwarz, JHEP. 0809, 027 (2008)
 \bibitem{ab1}
 M. Schnabl and Y.  Tachikawa, JHEP. 1009, 103 (2010)
 \bibitem{ab0}A. Mauri, A. Santambrogio and S. Scoleri, JHEP. 04, 146 (2013) 
 \bibitem{mp}O. K.  Kwon, P. Oh and  J. Sohn, JHEP.  0908, 093 (2009)
 \bibitem{mp1}I. R. Klebanov and  G. Torri, Int. J. Mod. Phys. A25, 332 (2010)
 \bibitem{mp2} O. K.  Kwon, P. Oh, C.  Sochichiu and  J.  Sohn, JHEP.  1003,  092 (2010)
 \bibitem{mp0}H. Samtleben and  R.  Wimmer,  JHEP. 1010, 080 (2010) 
 \bibitem{5bc}
O. Aharony, O. Bergman and D. L. Jafferis,  JHEP. 0811, 043
(2008) 
\bibitem{5ba}
S. Cremonesi, JHEP. 1101, 076 (2011)  
\bibitem{5a1}
J. Evslin and S. Kuperstein, JHEP. 0912, 016 (2009)
\bibitem{5a2}
 J. A. Minahan, O. Ohlsson Sax and C. Sieg, J. Phys. A43,  275402 (2010) 
\bibitem{5a}
P. Caputa, C. Kristjansen and K. Zoubos, Phys. Lett. B 677, 197 (2009) 
\bibitem{1}O. Aharony, O. Bergman, D. L. Jafferis and J. Maldacena, JHEP 0810, 091  (2008) 
\bibitem{2}O. Aharony, O. Bergman and D. L. Jafferis, JHEP. 0811,  043  (2008)
\bibitem{3}W. J. Rey and J. T. Yee, Eur. Phys. J. C 22, 379 (2001)
\bibitem{4}J. M. Maldacena, Phys. Rev. Lett. 80, 4859 (1998)
\bibitem{5}N. Drukker and D. Trancanelli, JHEP. 1002, 058 (2010)
\bibitem{6}N. Drukker, J. Plefka and D. Young, JHEP. 0811, 019 (2008)
\bibitem{7}S. J. Rey, T. Suyama and S. Yamaguchi, JHEP. 0903, 127 (2009) 
\bibitem{7a}A. Kapustin, B. Willett and I. Yaakov, JHEP. 1003, 089 (2010) 
\bibitem{p1} A. M. Polyakov, Nucl. Phys. 164, 171 (1980)
\bibitem{pq1}H. M. Chan, J. Faridani and S. T. Tsou,  Phys. Rev. D52,
  6134 (1995) 
\bibitem{pq2} H. M. Chan, J. Faridani and S. T. Tsou, Phys. Rev. D53 7293 (1996)
\bibitem{1p}H. M. Chan  and S. T. Tsou, Phys. Rev. D56, 3646  (1997) 
\bibitem{pq4}H. M. Chan   J. Bordes, and S. T. Tsou, Int. Jour. Mod. Phys. A14, 2173 (1999)
\bibitem{p9}H. M. Chan  and S. T. Tsou, Acta.  Phys. Polon. B28, 3027 (1997) 
\bibitem{pq0}H. M. Chan  and S. T. Tsou, Acta. Phys. Polon. B33, 4041 (2002)
\bibitem{q01}H. M. Chan,  Int. J. Mod. Phys. A16, 163 (2001) 
\bibitem{qp1}H. M. Chan and S. T. Tsou,  Phys. Rev. D57, 2507  (1998) 
\bibitem{pq01}H. M. Chan  and S. T. Tsou, Acta. Phys. Polon. B28, 3041 (1997) 
\bibitem{pq5}H. M. Chan, J. Faridani, J. Pfaudler and S. T. Tsou, Phys. Rev. D55, 5015 (1997) 
\bibitem{pq}M. Faizal, Euro. Phys. Lett.  103, 21003 (2013) 
\bibitem{p2}
H. M. Chan, P. Scharbach and S. T. Tsou, Ann. Phys.
167 454 (1986) 
\bibitem{p3}
H. M. Chan  and S. T. Tsou, Act. Phys. Pol. B17, 259 (1986)
\bibitem{p4}
H. M. Chan  and S. T. Tsou, Some Elementary Gauge Theory Concepts, World Scientific, (1993)
\bibitem{abcd}V. Cardinali, L. Griguolo, G. Martelloni and D. Seminara, arXiv:1209.4032
\bibitem{zois} S. T. Tsou and I. Zois, Rept. Math. Phys. 45, 229 (2000)
\bibitem{d2} S. Mukhi and C. Papageorgakis, JHEP. 0805, 085 (2008) 
\bibitem{sxsw} T. Li, Y. Liu and D. Xie, Int. J. Mod. Phys. A24, 3039 (2009) 
\bibitem{xswd}   Y. Pang and T. Wang, Phys. Rev. D78, 125007 (2008) 
\bibitem{d21}P. M. Ho, Y. Imamura and Y. Matsuo,  JHEP. 0807, 003 (2008) 
\bibitem{dual} J. Gomis and F. Passerini, JHEP. 0608, 074 (2006)
\bibitem{2dual}S. Yamaguchi, JHEP. 0605, 037 (2006)
\bibitem{dual2}J. Kluson and K. L. Panigrahi, Eur. Phys. J. C67, 565 (2010)
\bibitem{mf}M. Faizal and D. J. Smith, Phys. Rev. D 85, 105007 (2012)
\bibitem{mf2}D. S. Berman and D. C. Thompson, Nucl. Phys. B820, 503 (2009)
\bibitem{mf1}M. Faizal, JHEP. 1204, 017 (2012)
\bibitem{BCIntMem} D. S. Berman, M. J. Perry, E. Sezgin and D. C. Thompson, JHEP 1004, 025 (2010)
\bibitem{ChuSehmbi} C. S. Chu and G. S. Sehmbi, J. Phys. A 44, 134504 (2011)
\bibitem{d12}C. S. Chu and D. J. Smith,  JHEP. 0904, 097 (2009) 
\bibitem{M5BLG} P. M. Ho, Chin. J. Phys. 48, 1 (2010)
\bibitem{NCS1} E.~Bergshoeff, D.~S.~Berman, J.~P.~van der Schaar and P.~Sundell, Nucl. Phys. B 590, 173 (2000)
\bibitem{NCS2} S.~Kawamoto and N.~Sasakura, JHEP 0007, 014 (2000)
\bibitem{NCS3} D.~S.~Berman and B.~Pioline, Phys. Rev. D 70, 045007 (2004)
\end{thebibliography}
\end{document}